\documentclass[twocolumn]{autart}    

\usepackage{graphicx}          
\usepackage{amssymb}
\usepackage{xcolor}
\usepackage{palatino,epsfig,amssymb,cite,xcolor}
\usepackage[fleqn]{amsmath}
\usepackage{subfigure}
\usepackage{mathrsfs}  
\def\QED{~\rule[-1pt]{5pt}{5pt}\par}

\newcommand{\RL}{\mathcal{RL}}
\newcommand{\RH}{\mathcal{RH}}
\newcommand{\RHinf}{\mathcal{RH}_\infty}
\newcommand{\Linf}{L_\infty}
\newcommand{\Hinf}{H_\infty}
\newcommand{\T}{\mathbb{T}}
\newcommand{\D}{\mathbb{D}}
\newcommand{\enw}{\mathrm{e}^{-j\omega}}
\newcommand{\ew}{\mathrm{e}^{j\omega}}
\newcommand{\esw}{\mathrm{e}^{j2\omega}}

\newcommand{\net}{\mathrm{e}^{-\frac{T}{\tau}}}

\newcommand{\IC}{\mathbb{C}}
\newcommand{\cG}{\mathcal{G}}
\newcommand{\RF}{\mathbb{RF}}
\newcommand{\AP}{\mathbb{AP}}

\begin{document}

\begin{frontmatter}

\title{Exact Instability Radius of Discrete-Time {LTI} Systems} 

\thanks[footnoteinfo]{This work was supported by the National Science and Technology Council of Taiwan (grant numbers: 110-2221-E-110-047-MY3 and 113-2222-E-110-002-MY3).}

\author[Taiwan]{Chung-Yao Kao}\ead{cykao@mail.nsysu.edu.tw},         
\author[Taiwan]{Sei Zhen Khong}\ead{szkhong@mail.nsysu.edu.tw},      
\author[Japan]{Shinji Hara}\ead{shinji\_hara@ipc.i.u-tokyo.ac.jp},
\author[Taiwan]{Yu-Jen Lin}\ead{d083010007@student.nsysu.edu.tw}

\address[Taiwan]{Department of Electrical Engineering, National Sun Yat-sen University, 
          Kaohsiung 80424, Taiwan}  
\address[Japan]{Global Scientific Information and Computing Center, Tokyo Institute of Technology, Tokyo 152-8550, Japan}             
    

\begin{abstract}                          
The robust instability of an unstable plant subject to stable perturbations is of significant importance and arises in the study of sustained oscillatory phenomena in nonlinear systems. This paper analyzes the robust instability of linear discrete-time systems against stable perturbations via the notion of robust instability radius (RIR) as a measure of instability. We determine the exact RIR for certain unstable systems using small-gain type conditions by formulating the problem in terms of a phase change rate maximization subject to appropriate constraints at unique peak-gain frequencies, for which stable first-order all-pass functions are shown to be optimal. Two real-world applications --- minimum-effort sampled-data control of magnetic levitation systems and neural spike generations in the FitzHugh--Nagumo model subject to perturbations --- are provided to illustrate the utility of our results.

\end{abstract}

\end{frontmatter}

\section{Introduction}\label{intro}

Robust stability analysis, which assesses systems stability in the face of uncertainty and/or parametric variation,
is fundamentally important in the study of control systems~\cite{10.5555/225507}. The small-gain theorem is a 
classical robust stability result that quantifies the largest stable perturbation a stable system can tolerate 
before becoming unstable. This number is referred to as the ``robust stability radius'', which plays an essential 
role in certain approaches to designing feedback mechanisms, such as the $\mathcal{H}_{\infty}$ control~\cite{HINRICHSEN19861}.  
 
An important counterpart of the robust stability analysis is the so-called ``robust instability analysis'' for nominally unstable systems. 
Robust instability analysis considers the problem of quantifying the maximal stable perturbation that an unstable system can tolerate 
before becoming stable. Compared to the robust stability problem which has been extensively studied and is well understood, research on 
robust instability is not much less emphasized. However, the analysis does find important applications in, for example, feedback control 
to maintain nonlinear oscillation under the influence of structured perturbations. Robustness of such non-equilibrium states is difficult to analyze 
in general, and robust instability analysis (of linear systems) provides key insights for maintaining certain classes of nonlinear oscillations
as demonstrated in~\cite{hara2020robust,hara2022instability}, where robust oscillations in the sense of Yakubovich are guaranteed by instability and ultimate boundedness~\cite{pogromsky1999diffusion}. 

Robust instability analysis is a much more difficult problem than robust stability, of which the challenges can be understood as follows. Unlike ``robust stability radius'', 
to characterize ``robust instability radius'' (RIR) one needs to find the smallest (stable) perturbation that moves not one but \emph{all} 
unstable poles to the stable region of the complex plane. As such, the small-gain condition in terms of the $\mathcal{L}_\infty$-norm 
is only sufficient but not necessary for the robust instability. Furthermore, the search for the worst case stable perturbation in robust
instability analysis provides a solution to the strong stabilization; i.e., stabilization by a stable controller~\cite{zeren2000strong,Ohta1990ASO}. In fact, 
robust instability is equivalent to strong stabilization by a minimum-norm controller, which is a long-standing problem where challenges arise from its 
non-convexity nature and the fact that no upper bound on the order of stable stabilizing controllers has been found. It is noteworthy that the recent work~\cite{khong2024robust} examines the problem of robust instability from a different perspective involving the $\mathscr{L}_2$-gap.

Recently, the authors of~\cite{hara2023exact} considered the robust instability problem for classes of linear time-invariant (LTI) 
continuous-time systems which 
have one or two
unstable poles and a single peak-gain frequency, and provided an almost complete solution in terms of the small-gain-type condition. 
It was 
found that for the aforementioned classes of systems, the worst-case perturbations for a given unstable nominal plant are
those which 
compensate the phase of the plant at the peak-gain frequency and make the loop's ``phase change rate'' positive at that frequency. 
The discovery led to the so-called ``phase change rate (PCR) maximization'' problem for finding conditions 
to characterize robustness
instability radius. It is shown in~\cite{hara2023exact} that the solutions to the PCR maximization problem are provided by first-order all-pass functions. 
  
As a complement to the work of~\cite{hara2023exact}, this paper addresses the robust instability analysis of LTI discrete-time systems. In contrast to the 
preliminary work~\cite{lin2024exact}, which presents a concise framework for determining the exact RIR of discrete-time LTI systems, our objectives in this paper 
are to provide detailed proofs and offer several remarks on the main results to highlight noteworthy aspects. Additionally, we describe practical applications in 
neural dynamics and magnetic levitation to demonstrate how our findings offer valuable insights into real-world problems. The results in this paper are derived 
purely in the discrete-time setting and the mathematical expressions of the conditions, as well as the derivations of these results, 
are significantly distinct from their continuous-time counterparts. It is noteworthy that it is not viable to reproduce these results using the findings 
in~\cite{hara2023exact} by applying the bilinear transformation to discrete-time transfer functions. This is because the use of the bilinear transformation 
generally gives rise to biproper continuous-time transfer functions, whereas only strictly proper continuous-time transfer functions were investigated 
in~\cite{hara2023exact}.

The remainder of this paper is arranged as follows. Notation and terminology are defined in Section~\ref{def}. 
In Section~\ref{marginal} we present a necessary and sufficient condition for marginal stabilization of certain classes of LTI discrete-time systems. This is applied to derive conditions for such systems 
to have exact RIR in Section~\ref{result}. Similar to the continuous-time setting, the proof of the main results in Section~\ref{result} 
relies on solving a phase change rate maximization problem, which is examined in Section~\ref{sup}. Two examples demonstrating our results
are given in Section~\ref{ex} and some concluding remarks are drawn in Section~\ref{conclusion}. 

\section{Notation and Terminology}\label{def}
All systems considered in this paper are single-input-single-output, and therefore their frequency-domain representations are scalar transfer functions.
The unit circle in the complex plane $\IC$ is denoted by $\T$, while the open (closed) unit disk is denoted by $\D$ ($\overline{\D}$). 
We use $\RL_{\infty}$ to denote the set of all real rational functions that are bounded on $\T$. The subset of $\RL_{\infty}$ consisting of functions that are analytical and uniformly bounded outside $\overline{\D}$ is denoted by $\RH_{\infty}$. The norms in $\RL_{\infty}$ and 
$\RH_{\infty}$ are denoted by
$\| \cdot \|_{\Linf}$ and $\| \cdot \|_{\Hinf}$, respectively. Given $g\in \RL_{\infty}$ and $f\in\RHinf$, we define $\|g\|_{\Linf}:=\sup_{\omega\in(-\pi,\pi]}|g(\ew)|$ and $\|f\|_{\Hinf}:=\sup_{s\in\IC\backslash\overline{\D}}|f(s)|$. It can be shown that 
 $\|f\|_{\Hinf}=\sup_{\omega\in(-\pi,\pi]}|f(\ew)|$.

It is well-known that any discrete time, finite dimensional, linear time invariant and stable system can be identified with a frequency-domain transfer function in $\RH_{\infty}$. Consider the class of unstable systems that identify with transfer functions 
from the following set: $\mathcal{G}:=\{ g \in \RL_{\infty}\,|\, g \, \text{ is proper and has poles outside $\overline{\D}$}\}$. 
The robust instability radius for $g\in\mathcal{G}$ is defined as follows:
\begin{align*}
    \rho_*(g):=\inf_{\delta\in{\mathbb S}(g)}||\delta||_{\Hinf},
\end{align*}
where $\mathbb{S}(g)$ is the set of real-rational, proper, stable transfer functions that internally stabilize $g$. It is 
well-known that $\rho_*(g)$ is finite if and only if $g$ satisfies the so-called parity interlacing property (PIP) 
condition~\cite{youla1974single}, i.e., the number of unstable real poles between any pair of unstable real zeros is even, wherein strictly proper systems are considered to have one real unstable zero at infinity.
As such, the class of systems of our interest
is
\begin{align*}
{\mathcal G}_n:=\{g\in {\mathcal G}\,|\,g\,&\text{ has $n$ unstable poles and} \\ 
&\text{ satisfies the PIP condition}\}.
\end{align*} 
It is known that the $\RL_{\infty}$-norm of $g$ serves as a lower bound of $\rho_*(g)$; i.e., $\|g\|_{\Linf}\le\rho_*(g)$.
A system $g$ 
is said to have ``exact RIR'' if $\rho_*(g)$ is exactly equal to $\|g\|_{\Linf}$. Our primary interest here is to derive 
conditions on $g\in\mathcal{G}_n$ under which $g$ has exact RIR.
To this end, let us 
introduce the following notion of \emph{marginal stability} to facilitate a clear presentation of our results. 

A discrete-time system with transfer function $g$ is called \emph{marginally stable} if all its poles are in $\overline{\D}$, and those on $\T$ are simple. A marginally stable system $g$ is called ``\emph{single-mode}'' marginally stable if all its poles 
are in $\D$ except for a complex conjugate pair of poles on $\T$, one pole at $1$, or one pole at $-1$. 

In the continuous-time setting, it is shown in~\cite{hara2022instability} that, given an unstable system $g$, if 
a stable system $\delta$ has no unstable pole/zero cancellation with $g$, and the positive feedback system with loop transfer 
function $\delta g$ is single-mode marginally stable, then the $\mathcal{H}_\infty$-norm of $\delta$ provides an upper bound
for $\rho_*(g)$. The conclusion can be reached in the discrete-time setting via the bilinear transformation, which is known to preserve stability. 
As such, in this paper the same technical mechanism via single-mode marginal stabilization is adopted for obtaining 
conditions for ``exact RIR''.

\section{Marginal Stabilization of Discrete-Time LTI Systems with a Single Peak-gain Frequency}
\label{marginal}
An extended version of the Nyquist criterion for characterizing single-mode marginal stability of discrete-time LTI systems is 
presented in this section. It will be used for deriving stabilizing control synthesis conditions and conditions for characterizing systems with 
exact RIR. The results presented here are the discrete-time counterparts to those in~\cite{hara2023exact}. To this end, let us first introduce the concept 
of magnitude and phase change rates for an $\RL_\infty$ function. 

Let $L$ be a complex function from the class $\RL_\infty$. The logarithmic gain and the phase of $L$ at a complex number $z$
such that $L(z)\not=0$ are denoted by $\ln|L(z)|$ and $\angle L(z)$, respectively. Note that $\angle L(z)$ should be chosen so that $\omega \mapsto \angle L(\ew)$ is continuous on $(-\pi,\pi]$, except where $L(\ew) = 0$. Denote the logarithmic gain and the phase of the frequency
response $L(\ew)$ by
\begin{align*}
    A_L(\omega):=\ln|L(\ew)|\,,\quad \theta_L(\omega):=\angle L(\ew).
\end{align*}
Define the gain and the phase change rate of $L$ w.r.t. $\omega$ as
\begin{align*}
A_L'(\omega):=\frac{d}{d\omega}\ln|L(\ew)|\,,\quad \theta_L'(\omega):=\frac{d}{d\omega}\angle L(\ew).
\end{align*} 
For $L\in\RL_\infty$ with no zeros on $\T$, $\log L(\ew)$ is well-defined as $A_L(\omega)+j\cdot \theta_L(\omega)$. 
Therefore, $A_L'(\omega)$ and $\theta_L'(\omega)$ can be found by computing the real and imaginary parts of 
$\frac{d}{d\omega}\log L(\ew)$, respectively. 

\subsection{Criterion for marginal stability}
We provide conditions for marginal stability of a given positive feedback
system with an unstable loop transfer function $L$.
To facilitate the development, let us define the following quantities. Consider
the Nyquist plot of $\displaystyle \left. L(z^{-1})\right|_{z=(1-\epsilon)\ew}$, where $\epsilon \ge 0$ 
and $\omega$ ranges from $-\pi$ to $\pi$. 
Define $\nu_{+}(\epsilon)$ as the number of transverse crossing points on the real semi-interval $(1,\infty)$ 
from the negative imaginary region to the positive one by the Nyquist plot, and similarly $\nu_{-}(\epsilon)$
from positive to negative. Let $\nu_o(\epsilon):=\nu_{+}(\epsilon)-\nu_{-}(\epsilon)$. 

The following lemma states the discrete-time counterpart of Lemma 4 of~\cite{hara2023exact}, which provides a necessary and 
sufficient condition for marginal stability. Note that the statement, which makes use of the modified Nyquist contour 
$\mathcal{N}_{\epsilon^-}$ is $\{z~|~z=(1-\epsilon)\ew, \ \omega\in[-\pi,\pi], \epsilon\ge 0 \}$, is essentially an extended version of the classical 
Nyquist criterion equipped with the standard Nyquist contour $\mathcal{N}_{0^-}$.
Notice the orientation of $\mathcal{N}_{\epsilon^-}$ is counter-clockwise. 

\begin{lem}\label{Nyquist}
Consider a positive feedback discrete-time system with loop transfer function $L\in\mathcal{G}_n$. The feedback system has all poles in 
$\overline{\D}$ if and only if there exists an $\epsilon_{+}>0$ such that the following two equivalent conditions hold for all 
$\epsilon\in(0,\epsilon_{+})$.
\begin{enumerate}
    \item[(1)] 
    The number of clockwise encirclements of the Nyquist plot of 
    $\displaystyle \left. L(z^{-1})\right|_{z\in\mathcal{N}_{\epsilon^-}}$ about $1+j0$ 
    is equal to $n$, the number of unstable poles of $L$.
    \item[(2)] $\nu_o(\epsilon)=-n$ for the Nyquist plot of $\displaystyle \left. L(z^{-1})\right|_{z\in\mathcal{N}_{\epsilon^-}}$
\end{enumerate}
Moreover, the feedback system is marginally stable if and only if the following two conditions hold in addition
to condition (1) or (2): (3-a) $\exists \,\omega$ such that $L(\ew)=1$; (3-b) If $L(\ew)=1$, then 
$\left.\frac{d}{dz}L(z)\right.|_{z=\ew}\not = 0$.
\end{lem}

\begin{pf}
First, we note that one can readily verify that $\nu_o(\epsilon)=-n$ means exactly that the 
number of clockwise encirclements of the Nyquist plot of $\displaystyle \left. L(z^{-1})\right|_{z\in\mathcal{N}_{\epsilon^-}}$ about 
$1+j0$ is equal to $n$. 

A proof for condition (1) can be derived from the classical Nyquist criterion. Let $\tilde{L}(z):=L(z^{-1})$
and note that an unstable pole of $L$ gives a pole of $\tilde{L}$ in $\D$. 
Moreover, let 
\[
G(z):=\frac{L(z)}{1-L(z)}, \quad \tilde{G}(z):=\frac{\tilde{L}(z)}{1-\tilde{L}(z)}.
\]
Clearly $\tilde{G}(z)=G(z^{-1})$, and therefore $G$ has all poles in $\overline{\D}$ if and only if $\tilde{G}$ 
has no pole in $\D$. To show the latter, note that the Nyquist contour $\mathcal{N}_{\epsilon^-}$ is a counter-clockwise 
oriented circle with radius smaller than and arbitrarily close to 1 as $\epsilon\to 0$. Condition (1) implies that
the number of counter-clockwise encirclements  of the Nyquist plot of 
$\displaystyle \left. \tilde{L}(z)\right|_{z\in\mathcal{N}_{\epsilon^-}}$ about $1+j0$ is $-n$, matching the number of the 
poles of $\tilde{L}$ in $\D$. Therefore, by the Nyquist criterion, the closed-loop system $\tilde{G}$ has no poles in $\D$,
which in turn implies that all poles of $G$ are in $\overline{\D}$. Finally, the additional conditions (3-a) and (3-b) 
is equivalent to that $G$ has poles on $\T$ and that those poles are simple. By definition this means $G$ is marginally
stable. \hfill\QED
\end{pf}

\begin{rem}
Note that condition (1) is equivalent to that the number of \emph{counter-clockwise} encirclements of the Nyquist plot of 
$\displaystyle \left. L(z)\right|_{z\in\mathcal{N}_{\epsilon^+}}$ about $1+j0$ is equal to $n$, where the Nyquist contour 
$\mathcal{N}_{\epsilon^+}$ is defined exactly the same as $\mathcal{N}_{\epsilon^-}$ except that $(1-\epsilon)$ is 
replaced by $(1+\epsilon)$. Proposition~\ref{MstableViaPcr} below is derived using the equivalent condition. 
\end{rem}

Built upon Lemma~\ref{Nyquist}, the following proposition gives a necessary and sufficient condition for single-mode marginal stability. 
\begin{prop}\label{MstableViaPcr}
Consider a positive feedback system with loop transfer function $L\in\mathcal{G}_n$, where $n$ is a positive integer. Suppose that
$\omega_c\in[0,\pi]$ is given such that $A_L'(\omega_c)=0$. The feedback system is single mode marginally stable if and only if condition (i) 
and one of conditions (ii-a) and (ii-b), indicated below, are satisfied.
\begin{enumerate}
    \item [(i)] $L$ satisfies the followings:
    \begin{align*}
        &L(\mathrm{e}^{j\omega_c})=1,\quad \left.\frac{d}{dz}L(z)\right|_{z=\mathrm{e}^{j\omega_c}}\neq 0\\
        &L(\mathrm{e}^{j\omega_c})\neq 1,\quad \forall \omega \neq \pm \omega_c.
    \end{align*}
    \item [(ii-a)] The Nyquist plot of $\left.L(z)\right|_{z\in\mathcal{N}_0}$ satisfies either $\nu_o(0)=n-1$ if $\omega_c\in\{0,\pi\}$ or 
      $\nu_o(0)=n-2$ if $\omega_c\notin \{0,\pi\}$, and $ \theta_L'(\omega_c)>0$.   
    \item [(ii-b)] 
     The Nyquist plot of $\left.L(z)\right|_{z\in\mathcal{N}_0}$ satisfies $\nu_o(0)=n$ and $ \theta_L'(\omega_c)<0$. 
\end{enumerate}
\end{prop}
\begin{pf}
The result may be established using similar arguments to those of Proposition 2 in~\cite{hara2023exact}.  \hfill\QED
\end{pf}

\section{Main Results}\label{result}
In this section, we exploit Proposition~\ref{MstableViaPcr} to obtain conditions for (single mode) marginal 
stabilizability of certain classes of unstable discrete-time LTI systems. They are in turn used to derive conditions 
for those systems to have exact RIR. Specifically, we consider the following subsets of $\mathcal{G}_n$. Given $\omega_*\in[0,\pi]$, define
\begin{align*}
{\mathcal G}^{\omega_*}_{n} &:=\{g\in \cG_{n}~|~ \|g\|_{\Linf} =|g(\mathrm{e}^{j\omega_*})| >|g(\ew)|,  \\ 
& \hspace{3.75cm} \forall\,\omega\in[0,\pi], \ \omega\not=\omega_* \},
\end{align*}
${\mathcal G}^{0,\pi}_{n}:={\mathcal G}^{0}_{n}\cup{\mathcal G}^{\pi}_{n}$ and 
${\mathcal G}^{\#}_{n}:= \bigcup_{\omega_*\in(0,\pi)}{\mathcal G}^{\omega_*}_{n}$.
These subsets contain functions from $\cG_n$ which have an unique peak gain frequency. The assumption on the 
unique peak gain frequency is not unreasonable from the practical point of view; furthermore, extending our results to cover 
systems with multiple peak gain frequencies is possible in a similar vein to the continuous-time counterpart in~\cite{hara2023exact}.

\subsection{Conditions for marginal stability}
The following theorem provides necessary and sufficient conditions for marginal stabilization 
of system $g$ in ${\cG}_n^{0,\pi}$ or ${\cG}_n^{\#}$, by some system $f$ satisfying 
$\|f\|_{\Hinf}=1/\|g\|_{\Linf}$. Furthermore, the unique-peak-gain property of $g$
implies that the marginal stability is of single mode.   
As $1/\|g\|_{\Linf}$ is a lower bound on the RIR of $g$, single mode 
marginal stabilization of $g$ by such $f$ implies that $g$ has the exact RIR equal to $1/\|g\|_{\Linf}$,
as explained at the end of Section~\ref{def}.

\begin{thm}\label{condition4marginal}
The following hold on the classes ${\mathcal G}^{0,\pi}_{n}$ and ${\mathcal G}^{\#}_{n}$:
\begin{enumerate}
\item[(I)]
Let $g\in{\mathcal G}^{0,\pi}_{n}$ whose peak gain frequency $\omega_p$ is either $0$ or $\pi$. 
Then $g$ can be marginally stabilized by a stable system $f$ with $||f||_{\Hinf}=1/||g||_{\Linf}=1/|g(\mathrm{e}^{j\omega_p})|$ 
if and only if $n=1$ and 
\begin{align}
    \theta_g'(\omega_p)>0\ \label{eq:condmargin1}.
\end{align}
\item[(II)]
Let $g\in\mathcal{G}_n^{\#}$ whose peak gain occurs at $\omega_p$. Then $g$ can be marginally stabilized by 
a stable system $f$ with $||f||_{\Hinf}=1/||g||_{\Linf}=1/|g(\mathrm{e}^{j\omega_p})|$ if and only 
if $n=2$ and
\begin{align}
   \theta'_g(\omega_p)> |\sin(\theta_g(\omega_p))/\sin\omega_p |\label{eq:condmargin2}.
\end{align}
\end{enumerate}
\end{thm}
\begin{pf}
The proof follows arguments similar to those of Theorem~1 of~\cite{hara2023exact}. Here we briefly explain the underlying ideas but handle the differences with care. The sufficiency is proven by explicitly identifying $f$ which marginally stabilizes $g$ 
and applying Proposition~\ref{MstableViaPcr}. For $g\in\cG_n^{0,\pi}$, or $g\in\cG_n^{\#}$ such that
$\sin(\theta_g(\omega_p))=0$, the marginally stabilizing $f$ is a real constant whose magnitude equals to $1/\|g\|_{\Linf}$. For 
$g\in\cG_n^{\#}$ such that $\sin(\theta_g(\omega_p))\not=0$, $f$ is a first-order all-pass function of the form 
$\frac{1}{\|g\|_{\Linf}}\left(\frac{az+1}{z+a}\right)$, where $|a| < 1$ is chosen so that $L(z):=g(z)f(z)$ is equal 
to $1$ at 
$z=\mathrm{e}^{j\omega_p}$. By these choices of $f$, one can verify that the corresponding loop-gain $L$ satisfies 
conditions (i) and (ii-a)
of Proposition~\ref{MstableViaPcr}. Therefore, the positive feedback system with loop-gain $L:=gf$ is single mode 
marginally stable. 

For necessity, the arguments go as follows. Assume that there exists $f$ with 
$||f||_{\Hinf}=1/||g||_{\Linf}=1/|g(\mathrm{e}^{j\omega_p})|$ 
such that the the positive feedback system with loop-gain $L:=gf$ is marginally stable. Note that in this case, $|L(\ew)|$ 
is less than or equal to $1$ for all $\omega\in (-\pi,\pi]$. This property in turn implies that the marginal stability 
must be of single mode, and $\nu_o(0)$ must be equal
to $0$. Thus, conditions (i) and (ii-a) of Proposition~\ref{MstableViaPcr} must hold, and we have $n=1$ if peak gain 
of $g$ occurs at $\omega_p=0$
or $\pi$, and $n=2$ otherwise. Moreover, the condition $\theta_L'(\omega_p)>0$ implies 
$\theta_g'(\omega_p)>-\theta_f'(\omega_p)$, and 
the phase change rate conditions stated in~\eqref{eq:condmargin1} and~\eqref{eq:condmargin2} emerge as 
the infimum of $-\theta_f'(\omega_p)$, or equivalently, the supremum of $\theta_f'(\omega_p)$ over the following 
set of constraints: $f \in \RH_{\infty}$,
\begin{align}
\begin{split}\label{pcr_constraint}
&g(\mathrm{e}^{j\omega_p})f(\mathrm{e}^{j\omega_p})=1, \quad \text{and}\\
&|f(\ew)| \le |f(\mathrm{e}^{j\omega_p})| = 1/\|g\|_{\Linf}, \ \ \forall\omega\in [0,\pi].
\end{split}
\end{align}
We refer to such problem as ``phase change rate maximization'' problem, of which the solution is presented in Section~\ref{sup}. 
\hfill\QED
\end{pf}

The following remarks provide further insights about Theorem~\ref{condition4marginal}. 

\begin{rem}
We note important differences between the discrete-time result and its continuous-time counterpart. In 
the discrete-time setting, we allow for the scenario where the peak-gain frequency $\omega_p=\pi$; 
the corresponding continuous-time frequency would be $\omega_p=\infty$, which was not considered in~\cite{hara2023exact}.
In particular, the continuous-time setting in~\cite{hara2023exact} requires the system be strictly proper; as such, $\omega_p$
is always finite. Here, no such restriction on properness of the transfer function under study is imposed. Moreover, for 
continuous-time systems in $\cG_2^{\#}$, the corresponding
inequality~\eqref{eq:condmargin2} was found to be $\theta'_g(\omega_p)> |\sin(\theta_g(\omega_p))/\omega_p |$ ---
the denominator is $\omega_p$ instead of $\sin\omega_p$. This important difference can only be identified following arduous mathematical derivations performed in Section~\ref{sup}.
\end{rem}


\begin{rem}\label{rem:gn>=3}
For $g \in{\mathcal G}^{0,\pi}_{n}$ or ${\mathcal G}^{\#}_{n}$ with $n\geq 3$, it is still possible to find a stable $f$ 
with $||f||_{\Hinf}=1/||g||_{\Linf}$ such that all closed-loop poles are on $\overline{\D}$. However, 
the norm constraint on $f$ implies that the loop transfer function $L$ has a single peak at $\omega_p$, which in turn 
implies that the closed-loop poles on $\T$ must be located at $\pm\mathrm{e}^{j\omega_p}$, $1$ (when $\omega_p=0$), 
or $-1$ (when $\omega_p=\pi$). As such, the multiplicity of these poles must be larger than one, i.e. marginal stability does not hold.
\end{rem}

\subsection{Conditions for exact RIR}

We obtain here conditions
for a discrete-time system $g\in\cG_1^{0,\pi}$ or $\cG_2^{\#}$ to have exact RIR based on Theorem~\ref{condition4marginal}. 
Moreover, we also show that systems from the class of $\cG_1^{\#}$ do not have exact RIR.

\begin{lem}\label{necessaryRIR}
Given $\omega_p\in [0,\pi]$, integer $n\geq 1$, and $L\in\mathcal{G}_n$, 
consider the positive feedback system with loop transfer function $L$ satisfying the following conditions
\begin{align}
\hspace{-0.75cm}
1=\|L\|_{\Linf}=|L(\mathrm{e}^{j\omega_p})|>|L(\ew)|\,,\ 
\forall \omega\in[0,\pi]\backslash\{\omega_p\}
\label{eq:pg2}.
\end{align}
If the feedback system has all its poles in $\overline{\D}$, then $\theta'_L(\omega_p)\geq 0$.
\end{lem}
\begin{pf}
The arguments for proving Lemma~\ref{necessaryRIR} are identical to those of its continuous-time
counterpart~\cite[Lemma 5]{hara2023exact}. \hfill\QED
\end{pf}

\begin{thm}\label{Condition4RIR}
Let $g\in\cG$, and suppose $g(\ew)$ has a unique peak gain at $\omega_p$. Consider the exact RIR condition
\begin{align}
\label{cond_RIR}
\rho_*(g)=1/\|g\|_{\Linf}=1/|g(\mathrm{e}^{j\omega_p})|.
\end{align}
\begin{enumerate}
\item[(I)]
Suppose $g\in\cG_1^{0,\pi}$. Then 
\begin{align*}
\theta_g'(\omega_p)>0 \ \Rightarrow \ \eqref{cond_RIR} \ \Rightarrow \ \theta_g'(\omega_p)\ge 0.
\end{align*}
\item[(II)]
Suppose $g\in\cG_2^{\#}$. Then 
\begin{align*}
\theta_g'(\omega_p)>\varrho(\omega_p) \ \Rightarrow \ \eqref{cond_RIR} \ \Rightarrow \ \theta_g'(\omega_p)\ge \varrho(\omega_p), 
\end{align*}
where $\varrho(\omega_p) = |\sin\theta_g(\omega_p) / \sin\omega_p|$.  
\item[(III)] For any $g\in\cG_1^{\#}$, we have $\rho_*(g)>1/\|g\|_{\Linf}$.
\end{enumerate}
\end{thm}

\begin{pf} 
The sufficiency parts of the exact RIR conditions in statements (I) and (II) follow from
Theorem~\ref{MstableViaPcr} and the fact that if a stable function $f$ 
single-mode marginally stabilizes $g$, then one can find a stable $f_1$ 
such that $f+\varepsilon f_1$ with arbitrarily small $\varepsilon > 0$  stabilizes 
$g$~\cite[Proposition 1]{hara2023exact}. As such, $\rho_*(g)$ is equal to $\|f\|_{\Hinf}$. 

For the necessity part of the exact RIR condition in statement (I), suppose \eqref{cond_RIR} 
holds. This means one can find a stable $f$ such that $\|f\|_{\Hinf}=1/\|g\|_{\Linf}$ and the positive feedback system with loop gain $L:=gf$ has all its poles in $\overline{\D}$. 
One can verify that such $L$ would satisfy~\eqref{eq:pg2}, and therefore by Lemma~\ref{necessaryRIR} 
one has $\theta_L'(\omega_p)\ge 0$. This implies $\theta_g'(\omega_p)\ge -\theta_f'(\omega_p)$, 
which in turn leads to the inequality $\theta_g'(\omega_p)\ge 0$, since the infimum of $-\theta_f'(\omega_p)$ is equal to $0$ when $\omega_p = 0$ or $\pi$ as we will show in 
Section~\ref{sup}. 
The proof for the necessity part of the exact RIR condition in statement (II) follows exactly the same arguments. 
The only difference is that the infimum of $-\theta_f'(\omega_p)$ is equal to 
$|\sin\theta_f(\omega_p)/\sin\omega_p|$ when $\omega_p\in(0,\pi)$. 

For statement (III), consider a stable $f$ and the positive feedback 
system with loop gain $k\cdot L$, where $k\in(0,1]$ and $L:=gf$. Suppose that all poles 
of the closed-loop system are in $\overline{\D}$ when $k=1$. Since $L$ is real rational and has only one unstable pole, this pole must be on $(1,\infty)\cup(-\infty,-1)$ of the 
real axis, and there must be a $k\in(0,1]$ such that the closed-loop system has at least one pole at $1+j0$ or $-1+j0$, depending
on the location of the unstable pole in question. That is to say,  
$1 = k\cdot g(1)f(1)$, or $1=k\cdot g(-1)f(-1)$, for some $k\in(0,1]$. 
This implies $|f(1)| = 1/(k\cdot |g(1)|) \ge 1/|g(1)| > 1/\|g\|_{L_\infty}$, or 
$|f(-1)| = 1/(k\cdot |g(-1)|) \ge 1/|g(-1)| > 1/\|g\|_{L_\infty}$. 
The last inequality follows from the fact that $|g(\pm 1)|<|g(\mathrm{e}^{j\omega_p})|:=\|g\|_{\Linf}$ (for some $\omega_p\in(0,\pi)$) 
since $g\in\mathcal{G}_n^{\#}$. This shows 
that, any stable $f$ which renders all closed-loop poles in $\overline{\D}$ must satisfy $\|f\|_{\Hinf}>1/\|g\|_{\Linf}$.
Hence, $\rho_*(g)>1/\|g\|_{\Linf}$. \hfill\QED
\end{pf}
\begin{rem}
It should be clear from the proof that, the necessary conditions for the exact RIR hold for
$g\in\cG_n^{0,\pi}$ or $\cG_n^{\#}$ with any positive integer $n$. 
\end{rem}
\begin{rem}
For $g\in\cG_1^{0,\pi}$ or $\cG_2^{\#}$, the gap between the necessary and sufficient conditions for $g$ having the exact RIR 
arises from the multiplicity of the closed-loop poles on $\T$. The sufficient condition is obtained via (single-mode) marginal
stabilizability of $g$; this is not necessary because the transition from instability to stability might occur when multiple
poles pass the stability boundary at the same location.
\end{rem}
\begin{rem}
It can be shown that statement (III) holds for all $g\in\cG_{n}^\#$ where $n$ is an odd positive integer. To see this,
note that $g$ has an odd number of unstable poles, and therefore at least one unstable pole must transition from
instability to stability via $\pm 1$. The remaining arguments are identical to those presented above.    
\end{rem}
\begin{rem}
For $g\in\cG_1$ or $\cG_2$ with multiple peak gain frequencies, it can be single-mode marginally stabilized
by some stable $f$ with $\|f\|_{\Hinf}=1/\|g\|_{\Linf}$, and hence has the exact RIR, if 
\begin{itemize} 
\item for $g\in\cG_1$, either $|g(1)|$ or $|g(-1)|$ equal to $\|g\|_{\Linf}$, and 
inequality~\eqref{eq:condmargin1} holds correspondingly; 
\item for $g\in\cG_2$, there exists one peak frequency where inequality~\eqref{eq:condmargin2} holds.  
\end{itemize}
For such $g$, one can apply an inverse notch filter which maintains the gain at the frequency where~\eqref{eq:condmargin1} 
or~\eqref{eq:condmargin2} holds and decreases the gain at all other frequencies by
a sufficiently small amount. This way, the filtered $g$ has a unique peak frequency for which 
Theorems~\ref{condition4marginal} and~\ref{Condition4RIR} become applicable.
\end{rem}

\section{Phase Change Rate Maximization}
\label{sup}
The stabilizability conditions stated in Theorem~\ref{MstableViaPcr} and conditions for exact RIR in Theorem~\ref{Condition4RIR} 
emerge from the following phase change rate (PCR) maximization problem: 
\begin{align*}
\sup~\theta_f'(\omega_p),\ \ \mbox{subj. to $f\in\RH_\infty$ and~\eqref{pcr_constraint}}.
\end{align*}
In this section, the problem is solved, and it will be shown that the
supremums are attained by zeroth-order or first-order all-pass functions.

\subsection{Problem reformulation and results}
\label{subsec:pf_and_results}
Note that the phase variation $\theta_f'(\omega)$ of a function $f$ is invariant to 
constant scaling on $f$. Therefore, we can without loss of generality assume that 
$|f(\mathrm{e}^{j\omega_p})|=1/|g(\mathrm{e}^{j\omega_p})| =1$. As such, 
the essential aspects of the constraints in~\eqref{pcr_constraint} are
$\theta_f(\omega_p)=-\theta_g(\omega_p)$ and that $f$ attains its $\RH_\infty$ norm at $\omega_p$. 
Thus, it is equivalent to consider the following PCR maximization problem: 
\begin{align}
\begin{split}
&\hspace{-0.25cm} \sup~\theta_f'(\omega_p),\mbox{ subj. to $f\in\RH_\infty$,}  \\
& \|f\|_{\Hinf} = |f(\mathrm{e}^{j\omega_p})|=1, \ \theta_f(\omega_p)=-\theta_g(\omega_p) 
\end{split}
\label{pcr:equiv}
\end{align}
To facilitate the development, let us replace $-\theta_g(\omega_p)$ with $\theta_p$ for notational simplicity 
and consider
\begin{align*}
&\RF_{\omega_p,\theta_p}:=\{f\in\RH_\infty~|~\theta_f(\omega_p)=\theta_p, \\
&\hspace{3.5cm} \|f\|_{\Hinf} = |f(\mathrm{e}^{j\omega_p})|=1 \}\\
&\AP_{\omega_p,\theta_p}:=\AP\cap\RF_{\omega_p,\theta_p},
\end{align*}
where $\AP$ denotes the set of stable all-pass functions with unit gain. The optimization problem~\eqref{pcr:equiv}
is thus equivalent to 
\begin{align}
\sup_{f\in\RF_{\omega_p,\theta_p}} \ \theta_f'(\omega_p)
\label{opt_over_rf}
\end{align}
(with $-\theta_g'(\omega_p)$ replaced by $\theta_p$). The solution to this PCR maximization problem is stated below.

\begin{thm}
\label{thm:sol2pcr}
Consider the PCR optimization problem in~\eqref{pcr:equiv}  (with $-\theta_g'(\omega_p)$ replaced by $\theta_p$).
\begin{enumerate}
\item [(I)] For $\omega_p \notin \{0,\pi\}$ and $\theta_p\in (-\pi,\pi]$ (mod $2\pi$), we have
\begin{align*}
\hspace{-0.5cm}
\sup_{f\in \RF_{\omega_p,\theta_p}} \theta'_f(\omega_p) 
=-\left|\frac{\sin\theta_p}{\sin\omega_p}\right|.
\end{align*}
\item [(II)] For $\omega_p\in\{0,\pi\}$, whereby $\theta_p\in\{0,\pi\}$ (mod $2\pi$), it holds that
\begin{align*}
\hspace{-0.5cm}
\sup_{f\in\RF_{\omega_p,\theta_p}} \theta_f'(\omega_p)=0.
\end{align*}
\end{enumerate}
Moreover, when the supremum is zero (i.e., when $\theta_p=0$ or $\pi$), it is attained by either $f(z)=1$ or $f(z)=-1$. When the supremum is not zero, it is attained by a first-order all-pass 
function of the form $f(z)=\frac{az+1}{z+a}$, with $|a|<1$ chosen so that $\theta_f(\omega_p)=\theta_p$. 
\end{thm} 

Theorem~\ref{thm:sol2pcr} is a direct consequence of the following two propositions. Proposition~\ref{prop:pcr_red2AP}
shows that the solution of maximizing $\theta_f'(\omega_p)$ over $\RF_{\omega_p,\theta_p}$ is the same as that over 
$\AP_{\omega_p,\theta_p}$, and Proposition~\ref{prop:pcr_ap} gives the solution. 

\begin{prop}
\label{prop:pcr_ap}
The optimization problem
\begin{align}
\sup_{f\in \AP_{\omega_p,\theta_p}} \theta'_f(\omega_p)
\label{opt_over_ap}
\end{align}
admits the following solution:
\begin{enumerate}
\item [(I)] For $\omega_p \notin \{0,\pi\}$ and $\theta_p\in (-\pi,\pi]$ (mod $2\pi$), 
\begin{align*}
\hspace{-0.5cm}
\sup_{f\in \AP_{\omega_p,\theta_p}} \theta'_f(\omega_p) 
=-\left|\frac{\sin\theta_p}{\sin\omega_p}\right|
\end{align*}
\item [(II)] For $\omega_p\in\{0,\pi\}$, whereby $\theta_p\in\{0,\pi\}$ (mod $2\pi$),
\begin{align*}
\hspace{-0.5cm}
\sup_{f\in\AP_{\omega_p,\theta_p}} \theta_f'(\omega_p)=0.
\end{align*}
\end{enumerate}
The supremum is attained by either $f(z)=1$ or $f(z)=-1$ when it zero. 
When the supremum is not zero, it is attained by a first-order all-pass 
function of the form $f(z)=\frac{az+1}{z+a}$, with $|a|<1$ chosen so that $\theta_f(\omega_p)=\theta_p$. 
\end{prop}
\begin{prop}
\label{prop:pcr_red2AP}
We have
\begin{align}
\label{pcr:red2AP}
\sup_{f\in\RF_{\omega_p,\theta_p}}\theta_f'(\omega_p) = \sup_{f\in\AP_{\omega_p,\theta_p}}\theta_f'(\omega_p).
\end{align}
\end{prop}

The proof of Proposition~\ref{prop:pcr_ap} is given in Section~\ref{subsec:pf4prop2}, 
while that of Proposition~\ref{prop:pcr_red2AP} in Section~\ref{subsec:pf4prop3}. 
The results stated in Theorem~\ref{thm:sol2pcr} are similar to their continuous-time 
counterparts in Theorem~3 of~\cite{hara2023exact}. However, 
we note the following distinct feature of Theorem~\ref{thm:sol2pcr} compared to its 
continuous-time counterpart. When $\omega_p$ does not 
take the ``boundary'' values $\{0,\pi\}$, the supremum is $-|\sin\theta_p/\sin\omega_p|$
instead of $-|\sin\theta_p/\omega_p|$ in the continuous-time setting. On the other hand, 
since $\lim_{\omega_p\to\infty} -|\sin\theta_p/\omega_p| = 0$, the supremum at $\omega_p = \pi$ in the
discrete-time setting matches that in the continuous-time setting as the peak frequency approaches infinity.

\subsection{Proof of Proposition~\ref{prop:pcr_ap}}
\label{subsec:pf4prop2}

\begin{lem}
\label{lem:1stPCR}
Let $f(z)$ be a stable $n$th-order all-pass function with real poles (or no pole when $n=0$). Then for a given $\omega_p\notin\{0,\pm\pi\}$,
\begin{align}
\theta_f'(\omega_p) \le  -\left|\sin(\theta_f(\omega_p))/\sin\omega_p\right| . 
\label{eq:RateChangeReal_p1}
\end{align}
Moreover, when $n=0$ and $n=1$ the equality holds for all $\omega_p\notin\{0,\pm\pi\}$.\footnote{ 
When $n=0$, clearly $\theta_f'(\omega_p)=0$ for all $\omega_p\in[0,\pi]$.} 
\end{lem}

\begin{lem}  
\label{lem:c<r} 
Let $f_r$ and $f_c$ denote real rational $2$nd order stable all-pass functions with two real poles 
and a pair of complex conjugate poles, respectively.
For any $f_c$ and $\omega_p\notin\{0,\pm\pi\}$, there exists $f_r$ such that
\begin{align}
\theta_{f_c}(\omega_p) =\theta_{f_r}(\omega_p) \quad \mbox{and} \quad \theta_{f_c}'(\omega_p) < \theta_{f_r}'(\omega_p).\label{ineq:c<r}
\end{align}
\end{lem}

While the arguments of the proofs of Lemmas~\ref{lem:1stPCR} and~\ref{lem:c<r} are similar to their continuous-time counterparts in~\cite{hara2023exact}, the derivation of the mathematical expressions and technical details is laborious
and can be found in Appendices~\ref{appx:lem:1stPCR} and~\ref{appx:lem:c<r}, respectively. 

Now we are ready to prove Proposition~\ref{prop:pcr_ap}. 
Let $f$ be an $n$th-order real-rational all-pass function with unity norm. Then
$f$ can be factorized as $\pm f_1f_2\cdots f_m$, where $f_i$, $i=1,\cdots,m$ is either 
a first-order all-pass function, or a second-order one with a pair of complex conjugate
ploes, such that $f_i(1)=1$.  This follows from results on finite Blaschke products~\cite{Garcia2015}. 

First consider the case where $\omega_p\not\in\{0,\pi\}$. By Lemma~\ref{lem:c<r}, the optimal solution to 
the problem in~\eqref{opt_over_ap} would not have poles with nonzero imaginary parts; otherwise, one could replace these (2nd order) factors having 
complex poles with those having real poles, and the phase change rate would be elevated, which leads to a contradiction. 
As a result, the supremum of~\eqref{opt_over_ap} can be 
found by searching over the subset of $\AP_{\omega_p,\theta_{p}}$ 
containing real-rational all-pass functions which are constant functions or can be factorized
as products of first-order factors. Lemma~\ref{lem:1stPCR} then gives an upper bound on the supremum in question.
One can verify that the upper bound is attained by a first-order all-pass function $f(z)=\pm\frac{az+1}{z+a}$, with $|a|<1$ chosen such that
the constraint $\theta_f(\omega_p)=\theta_p$ is observed. On the other hand, if $\theta_p \in\{0,\pi\}$, then the upper bound is equal to
zero and attained by constant function $f(z)=1$ or $f(z)=-1$. 

Next, suppose $\omega_p\in\{0,\pi\}$. Since $f$ is real rational,  $\theta_f(\omega_p)$, which is constrained to be equal to $\theta_p$, 
must be either $0$ or $\pi$. In this case, it is clear that the constant function $f(z)=1$ or $f(z)=-1$ is the optimal solution, which has
zero PRC at all frequencies. To see this, note that any stable all-pass functions of order $n\ge 1$ can be factorized into the 
product of 1st- and/or 2nd-order components. One can readily verify that the PCRs of 1st- and 2nd-order all-pass functions at 
the frequency 0 and $\pi$ are strictly negative, which in turn implies that $\theta_f'(\omega_p)|_{\omega_p=0 \ {\rm or}  \ \pi} < 0$
for any stable all-pass $f$ of order larger than 0. 

\subsection{Proof of Proposition~\ref{prop:pcr_red2AP}}
\label{subsec:pf4prop3}

The main idea behind the proof of Proposition~\ref{prop:pcr_red2AP}
is described below. Given any $f\in\RF_{\omega_p,\theta_p}$, $f$ can
be factorized as $f_if_o$, where $f_i$ is an all-pass function, and $f_o$ is stable and minimum-phase. 
The key observation that leads to~\eqref{pcr:red2AP} is that the PCR of $f_o$ at the 
peak-gain frequency $\omega_p$ is always negative and renders no ``help'' in elevating 
the PCR of $f$. To show this result, we require the following technical lemma, whose proof is given 
in the Appendix~\ref{appx:mp<1st}. 

\begin{lem}
\label{mp<1st}
Given a minimum-phase system $f\in \RH_\infty$, suppose the peak-gain frequency $\omega_p\not\in\{0,\pm\pi\}$ and 
$\theta_f(\omega_p) \in (-\pi, \pi]$. Then
\begin{align*}
    \theta'_f(\omega_p)\leq -\bigg|\frac{\theta_f(\omega_p)}{\sin\omega_p}\bigg|
\end{align*}
Moreover $\theta_f'(\omega_p)\leq 0$ regardless of the value of the peak-gain frequency $\omega_p$. 
\end{lem}

To facilitate the development, let us define
\begin{align*}
\mathcal{O}_{\omega_p, \theta_p} := \{f & \in \RHinf : f \text{ is minimum-phase, }\\
 &|f(j\omega_p)| = \|f\|_{\Hinf}, \text{ and } \theta_f(\omega_p) = \theta_p\}.
\end{align*}
First consider the case where $\omega_p\not\in \{0,\pi\}$ and $\theta_p\in(-\pi,\pi]$. 
Since every $f \in \RHinf$ admits an inner-outer factorization $f = f_if_o$ with $f_i \in \RHinf$ 
being all-pass and $f_o \in \RHinf$ being minimum-phase, it follows that ${  |f_o(j\omega_p)|} = \|f_o\|_{\Hinf}$ and
  \begin{align*}
  &\hspace{-0.7cm} \sup_{f \in \RF_{\omega_p, \theta_p}} \theta_f^\prime (\omega_p) 
     = \sup_{f = f_if_o \in \RF_{\omega_p, \theta_p}} \left(  \theta_{f_i}^\prime (\omega_p) +
        \theta_{f_o}^\prime (\omega_p) \right) \\
    &\hspace{-0.7cm} \le \sup_{\theta \in (-\pi, \pi]} \left( \sup_{f_i \in \AP_{\omega_p, \theta}}  \theta_{f_i}^\prime
      (\omega_p) + \sup_{f_o \in {  \mathcal{O}_{\omega_p, \theta_p - \theta}} } \theta_{f_o}^\prime (\omega_p) \right) \\
    &\hspace{-0.7cm} \le \sup_{\theta \in (-\pi, \pi]} \left( -\frac{|\sin(\theta)|}{|\sin\omega_p|} -
      \frac{|\theta_p - \theta|}{|\sin\omega_p|} \right) 
    \le -\left|\frac{\sin(\theta_p)}{\sin\omega_p}\right|,
  \end{align*}
where the second inequality follows from Proposition~\ref{prop:pcr_ap} and Lemma~\ref{mp<1st}, and the last inequality
follows from the fact that $|b - a| + |\sin(a)| - |\sin(b)| \geq 0$ for all $a, b \in (-\pi, \pi]$.
Lastly, {  since $-\left|\sin(\theta_p)/\sin\omega_p\right| = 
\sup_{f \in \AP_{\omega_p, \theta_p}} \theta_f^\prime (\omega_p)  \le \sup_{f \in \RF_{\omega_p, \theta_p}} \theta_f^\prime (\omega_p)$, 
we conclude that the supremum of $\theta'(\omega_p)$ over $\RF_{\omega_p, \theta_p}$ is the same as that over $\AP_{\omega_p, \theta_p}$.}

Now consider the case where $\omega_p=0$ and $\theta_p\in\{0,\pi\}$. The reason why the value of 
$\theta_p$ is restricted to $0$ or $\pi$ is due to $f$ being real rational and therefore 
$\theta_f(\omega_p)\in\{0,\pi\}$. The rest of the derivation follows the same arguments. By the inner-outer 
  factorization of $f$, we have
  \begin{align*}
    &\sup_{f \in \RF_{0, \theta_p}} \theta_f^\prime (0) 
     = \sup_{f = f_if_o \in \RF_{0, \theta_p}} \left(  \theta_{f_i}^\prime (0) +
        \theta_{f_o}^\prime (0) \right) \\
    & \le \sup_{\theta \in \{0,\pi\}} \left( \sup_{f_i \in \AP_{0, \theta}}  \theta_{f_i}^\prime
      (0) + \sup_{f_o \in \mathcal{RF}_{0, \theta_p - \theta}} \theta_{f_o}^\prime (0) \right) \\
    & \le \sup_{\theta \in \{0,\pi\}} \sup_{f_i \in \AP_{0, \theta}}  \theta_{f_i}^\prime(0)=0.
  \end{align*}
The second inequality follows from Lemma~\ref{mp<1st}, where it is shown that $\theta_{f_o}'(\omega_p) \le 0$ 
for any minimum-phase $f_o\in\RHinf$. The last equality follows from Proposition~\ref{prop:pcr_ap}. This concludes the proof.

\section{Practical Applications}
\label{ex}
In this section, we apply our main results to real-world applications to demonstrate 
their practical relevance. The first application involves designing a minimum-norm stable
controller for a magnetic levitation system, which is within the class $\cG_1^{0,\pi}$. 
The second application involves analyzing neural dynamics in robust generation of 
action potentials via sustained oscillatory behavior. We focus on the second-order nonlinear 
FitzHugh-Nagumo (FHN) model~\cite{fitzhugh1961impulses,kwessi2024nearly}, of which the 
linearized approximation around its equilibrium falls within the class  $\cG_2^{\#}$.

\subsection{Strong stabilization of magnetic levitation systems}
A typical linearized model for the magnetic levitation system~\cite{namerikawa2001uncertainty} 
at an equilibrium is represented by
\begin{align}
\label{glev}
g(s) = k/\left((-s^2+p^2)(\tau s+1)\right),
\end{align} 
where $k>0$, and the pair of poles at $\pm p$ is due to the mechanical aspect of the system 
while the stable pole at $-\tau^{-1}$ arises from the electrical part. Here we consider 
minimum-norm strong stabilization of~\eqref{glev} in the digital control setting.
Assume that an ideal sampler and a synchronized zeroth-order hold with the sampling period 
$T>0$ are placed around the continuous-time plant $g(s)$. The time-discretized model is thus 
given by the following $Z$-domain transfer function 
\begin{align}
g_d(z) &= \frac{k}{p^2}\left(\frac{N_1}{z-\mathrm{e}^{pT}}+
\frac{N_2}{z-\mathrm{e}^{-pT}}+\frac{N_3}{z-\mathrm{e}^{-T/\tau}}\right)\notag \\
&= \frac{k}{p^2}\frac{\beta_2 z^2+\beta_1 z+\beta_0}{(z - \mathrm{e}^{pT})(z - \mathrm{e}^{-pT})
(z-\mathrm{e}^{-\frac{T}{\tau}})}, \label{eq:gz}
\end{align}
where $N_1:=\frac{-(\mathrm{e}^{pT}-1)}{2(\tau p+1)},\,
N_2:=\frac{-(1-\mathrm{e}^{-pT})}{2(\tau p-1)},\,
N_3:=\frac{\tau^2p^2(1-\mathrm{e}^{-T/\tau})}{\tau^2p^2-1}$, 
and $\beta_2:=N_1+N_2+N_3$, $\beta_1:=-(\e^{-pT}(N_1+N_3)+
\e^{pT}(N_2+N_3)+\e^{-T/\tau}(N_1+N_2))$, and
$\beta_0:=N_1\e^{-(pT+T/\tau)}+N_2\e^{(pT-T/\tau)}+N_3$.

Note that the static gain and the number of unstable poles are preserved;
i.e., $g_d(1) = g(0) = k/p^2$ and $g_d$ has one unstable pole.
Moreover, it can be verified by a lengthy calculation using certain 
preservation relations between real poles and zeros shown in~\cite{Hara1989_1,Hara1989_2} that 
the following features of $g$ are also preserved and inherited by $g_d$: 
$g_d\in\mathcal{G}_1^0$ and $\theta_{g_d}'(0) < 0$. By Theorem~\ref{Condition4RIR}, 
that $g_d$ has a strictly negative PCR at the peak-gain frequency implies that the 
necessary condition for $g_d$ to have the exact RIR is violated. Therefore, we 
conclude that 
\begin{align*}
   \rho_*(g_d):= \inf_{f\in\mathbb{S}(g_{d})}||f||_{\Hinf}> p^2/k.
\end{align*}
To find an upper bound on the RIR, let us introduce a stable high-pass function 
$f_h(z):=\frac{(b+1)z+1-b}{(a+1)z+1-a}$, $b>a>0$, 
which has a positive PCR at $\omega=0$. The idea is to apply $f_h$ so that the compensated 
plant $g_{dc}:=g_d f_h$ belongs to $\cG_1^{0}$ 
and satisfies $\theta_{g_{dc}}'(0)>0$. This way, we know $\rho_*(g_{dc})= 1/|g_{dc}(1)| = 
1/|g_d(1)| = p^2/k$. Furthermore, a
stable stabilizing controller $f$ for $g_{dc}$ would give a stable stabilizing controller $f 
f_h$ for $g_d$. As such, we conclude that 
\begin{align*}
\rho_*(g_d)\le \rho_*(g_{dc})\cdot\|f_h\|_{\Hinf}
=(b/a)(p^2/k).
\end{align*}
Moreover, to minimize the upper bound, we shall make $b/a$ as small as possible. 

Observe that 
\begin{align*}
 &A_{f_h}'(\omega)= 2(b^2-a^2)\sin\omega/\mathbf{D} \\
 &\theta_{f_h}'(\omega) = 2(b-a)\left((1-ab)+(1+ab)\cos\omega \right)/\mathbf{D},
\end{align*}
where $\mathbf{D}:=3a^2b^2+a^2+b^2+3+4(1-a^2b^2)\cos\omega+(1-a^2-b^2+a^2b^2)\cos2\omega$. 
To obtain $\theta_{g_{dc}}'(0)>0$, 
$\theta_{f_h}'(0)$ must be larger than $-\theta_{g_d}'(0)$; thus we require 
$b-a>-2\theta_{g_d}'(0)$. Let $b=a+\textbf{P}_\epsilon$, where $\textbf{P}_\epsilon:=-2\theta_{g_d}'(0)+\epsilon$ and $\epsilon$ is some positive real number. 

Note that $A_{g_{dc}}'(0)=0$ regardless of the choice of $a$ and $b$. To ensure that $g_{dc}\in\cG_1^{0}$, a sufficient condition is that $A_{g_{dc}}'(\omega)\leq 0,\,\forall \omega\in [0,\pi]$.
In light of the goal of minimizing the factor $b/a = 1+\textbf{P}_\epsilon/a$, we shall find the largest $a$ such that 
$A_{g_{dc}}'(\omega):=A_{g_d}'(\omega)+A_{f_h}'(\omega)$ is non-positive for all $\omega\in(0,\pi]$; i.e.,
find $\bar{a}=\max~a$, such that 
\begin{align*}
\hspace{-0.1cm}
\left.\frac{2(b^2-a^2)\sin\omega}{\mathbf{D}}\right|_{b=a+\mathbf{P}_\epsilon}\le
-A_{g_d}'(\omega) \quad \forall \omega\in[0,\pi].
\end{align*}
Note that $A_{g_d}'(\omega)$ is equal to the real part of $\frac{d}{d\omega}\log g_d(\ew)$, which can be readily computed for 
a given $g_d$. It can be verified that, for the $g_d$ under consideration, $\bar{a}$ is equal to
\begin{align*}
\hspace{-0.7cm}
&\frac{1}{2\textbf{P}_{\epsilon}}\left(\frac{8}{\mathrm{e}^{pT}+\mathrm{e}^{-pT}-2}+\frac{4\net}
{(1-\net)^2}+\right.\\
&\hspace{2.5cm}
\left.
\frac{4(4\beta_2\beta_0+\beta_2\beta_1+\beta_1\beta_0)}{(\beta_2+\beta_1+\beta_0)^2}-\textbf{P}_{\epsilon}^{2}\right),
\end{align*}
where the parameters $\beta_0,\ \beta_1, \ \beta_2$ are defined below~\eqref{eq:gz}. This leads to the following inequality:
\begin{align*}
1<\mathop\text{inf}\limits_{f\in\mathbb{S}(g_d)}\frac{||f||_{h_\infty}}{p^2/k}\le (1+\mathbf{P}_\epsilon/\bar{a}).
\end{align*}
To conclude this section, we make the following observations. When $\tau\to 0$, we have $\beta_0\to 0$, 
$\beta_2\to\beta_1\to (2-\e^{pT}-\e^{-pT})/2=:\kappa$, and $\mathbf{P}_{\epsilon}\to 1+\epsilon=:\mathbf{1}_\epsilon$, 
These in turn imply $\mathbf{P}_\epsilon/\bar{a}\to 2\mathbf{1}_\epsilon^2/(1-\frac{4}{\kappa}-\mathbf{1}_\epsilon^2)$. 
The upper bound is consistent with the one found in~\cite{lin2024exact}, where a reduced 2nd-order model for a magnetic levitation system was considered.
Furthermore, if in addition the sampling period $T\to 0$, we have $\kappa\to 0^-$ and $\mathbf{P}_\epsilon/\bar{a}\to 0^+$. 
The upper bound approaches the lower bound when the sampling period approaches 0, which is consistent with the result
reported in~\cite{kao2023phase}.


\subsection{Robustness of oscillatory behavior in neural dynamics}

We analyze the discrete-time FHN model from~\cite{kwessi2024nearly} subject to multiplicative perturbations. The FHN model describes an exitable system (e.g. neuron) that exhibits oscillatory behavior that is predicated on the instability of its equilibrium. Here, we determine the minimum amount of LTI multiplicative perturbation needed to render the equilibrium stable, which results in the cessation of the oscillation of the system. In a neuronal system, this would correspond to the breakdown of the function of neurotransmission.

The FHN model is represented by the following equations:
\begin{equation}\label{ex1:DFHT}
    \begin{aligned}
    &x_{n+1}=\frac{Ax_n+(1-A)((1+\delta(z))y_n-I)}{1+\frac{A-1}{3}x_n^2}\\
    &y_{n+1}=By_n+D(1-B)(x_n+\alpha),
\end{aligned}
\end{equation}
where $x$ represents the membrane potential, $y$  the recovery variable and $I$ the intensity of the injected current, assumed constant here. The term $\delta(z)y_n$ represents a time-domain signal obtained by the inverse Z-transform of the product of $\delta$ and Z-transform of $y$. Define $A:=\mathrm{e}^{c^{-1}\tau}$, $B:=\mathrm{e}^{\frac{-\beta}{d}\tau}$ and $D:=\beta^{-1}$, where $\alpha$ is the excitability parameter of the system, $\beta$ represents an excitation threshold, $c$ is a positive parameter determining the rate at which $x$ changes relatively to $y$, $d$ is a scaling parameter and $\tau$ is the step size. 

In the illustrative example below, the parametric values in the FHN model are given by $c=1$, $\alpha=0.7$, $\beta=0.8$, $\tau=0.01$, $d=10$, and $I=0.4$. Let $g$ represent the linearized model from~\eqref{ex1:DFHT} at the fixed point $(\bar{x}=-0.9066,\bar{y}=-0.2582)$ with perturbation $\delta(1)=0$ and $g_e$ the linearized model with perturbation $\delta(1)=e$, $e\in\mathbb{R}$. It can be verified that $g=\frac{(1.5679z-2.5685)10^{-5}}{z^2-2.000985z+1.000994}\in\cG_2^{\#}$ with $\rho_*(g)=0.2868$ using Theorem~\ref{Condition4RIR}. A complicating matter in subjecting $g$ to a perturbation $\delta$ such that $\delta(1) = e \ne 0$ is that the fixed point of~\eqref{ex1:DFHT} will undergo changes. Additionally, the originally unstable fixed point could become stable, leading to the existence of $|\delta(1)|<1/||g||_{\Linf}$ such that~\eqref{ex1:DFHT} is stable. In view of Theorem~\ref{thm:sol2pcr}, this $\delta$, which is not all-pass, is clearly not an optimal perturbation that achieves exact RIR of the FHN model.

\begin{figure}
    \centering
    \includegraphics[scale=0.35]{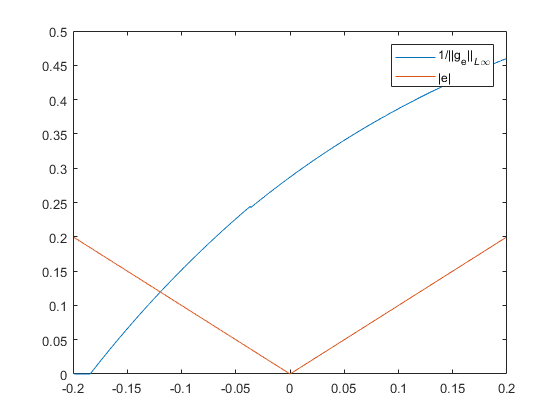}
    \caption{$\frac{1}{||g_e||_{\ell_\infty}}$ under different $e$}
    \label{fig:search_e}
\end{figure}

To determine the exact RIR for $g_{e}$ and hence the FHN model, we first search for $e$ such that $g_{e}$ satisfies the sufficient condition for exact RIR in Theorem~\ref{Condition4RIR} and then find the minimum value $e_o$ that satisfies $|e_o|=1/||g_{e_o}||_{\Linf}$. A numerical search as illustrated in Fig.~\ref{fig:search_e}. results in $e_o=-0.1192$, with the corresponding fixed point of the FHN model at $(\bar{x}=-0.9389,\,\bar{y}=-0.2986)$ and linearized dynamics of $g_{e_o}(z)=\cfrac{(1.8767z-2.8769)10^{-5}}{z^2-2.00039z+1.000399}$.
It can be verified that the unique peak gain of $g_{e_o}(\ew)$ occurs at $\omega_p=0.003$, indicating that $g_{e_o}\in\cG_2^{\#}$. Moreover, $\theta'_{g_{e_o}}(\omega_p)>|\sin\theta_{g_{e_o}}(\omega_p)/\sin\omega_p|$ holds, implying via Theorem~\ref{Condition4RIR} that the exact RIR for $g_{e_o}(z)$ is $\rho_*(g_{e_o})=|e_o|$. Furthermore, according to Theorem~\ref{thm:sol2pcr}, there exists a stable perturbation $\delta_f(z)$, which causes the solutions of characteristic equation $1-\delta_f(z)g_{e_o}(z)=0$ to be all in $\D$ except for a pair of complex poles on the unit circle. Such a stable perturbation is represented by a stable all-pass function $\delta_f(z)=0.1192\cdot\cfrac{0.9969z-1}{z-0.9969}$. 

\begin{figure}[t]
    \centering
    \subfigure[$\epsilon=-0.05$]{
    \includegraphics[scale=0.26]{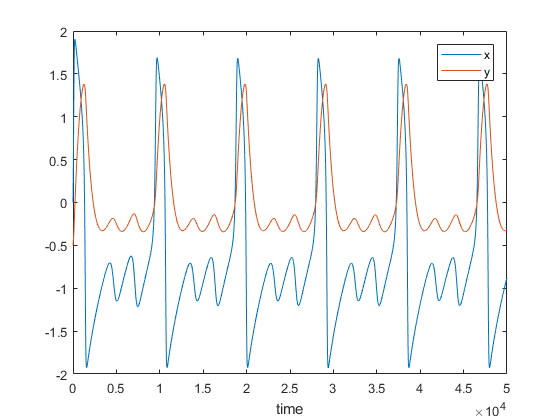}
    \label{fig:oscillation}
    }
    \subfigure[$\epsilon=0.05$]{
    \includegraphics[scale=0.26]{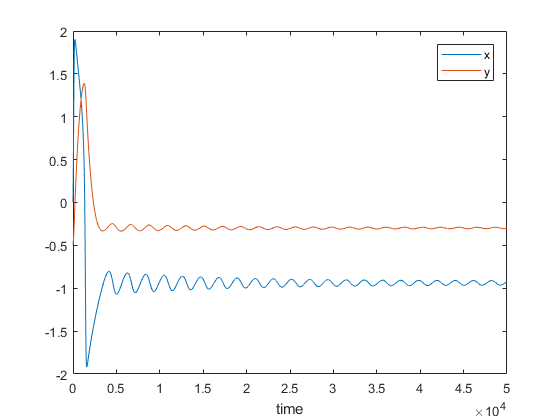}
    \label{fig:non-oscillation}}
    \caption{FHN response under $\delta_{f_\epsilon}$ with different $\epsilon$}
    \label{fig:trajectories}
\end{figure}

Next, we demonstrate via simulations that exact RIR of the FHN model has indeed been found. Let $\epsilon$ be a small real number and define $\delta_{f_\epsilon}(z):=(1+\epsilon)h(z)\delta_f(z)$, where $h(z)$ is a stable function such that $h(1)=\frac{1}{1+\epsilon}$ and $h(\mathrm{e}^{j\omega_p})=1$. Observe that $|h(1)| > |h(\mathrm{e}^{j\omega_p})|$ when $\epsilon<0$ and $|h(1)| < |h(\mathrm{e}^{j\omega_p})|$ when $\epsilon>0$. The purpose of introducing $h$ is to ensure that both $\delta_{f_\epsilon}(z)$ and $\delta_{f}(z)$ share the same DC gain, so that they give rise to the same fixed point of the FHN model when applied. It may be verified that the fixed point $(\bar{x},\bar{y})$ remains unstable when $\epsilon=-0.05$, whereas it becomes stable when $\epsilon=0.05$. Trajectories of the state $(x,y)$ with respect to the two different $\epsilon$ have been plotted in Fig.~\ref{fig:trajectories}. Evidently, the oscillatory phenomenon is sustained in the case where the fixed point is unstable, and suppressed otherwise.

\section{Conclusions}\label{conclusion}

In this paper, we established necessary and sufficient conditions for certain classes of unstable discrete-time 
plants to have the exact robust instability radius (RIR). These conditions are acquired via solving 
a phase change rate maximization problem, the solutions of which also lead to the minimum stable perturbation that destroys 
instability of an unstable plant, or from another point of view, the minimum stable control that stabilizes the plant. 
We demonstrated the utility of our results by two practical examples, including the strong stabilization of a magnetic levitation system 
and robust analysis of oscillatory behaviors in neural dynamics. 
The results in this paper provide the first step towards a complete theory of robust instability analysis. 
Extending the results to more general classes of unstable networked systems is currently being pursued.

\appendix

\section{Proof of Lemma~\ref{lem:1stPCR}}
\label{appx:lem:1stPCR}

As a consequence of the results on finite Blaschke products~\cite{Garcia2015}, 
an $n$th-order real-rational all-pass function $f$ with unit norm and only real poles
can be factorized as $f(z) = c f_1(z)...f_n(z)$, where 
$f_i(z)=\frac{a_iz+1}{z+a_i}$ with $|a_i|<1$, $i=1,\cdots,n$, and $c \in \{-1, 1\}$. 
Furthermore, since $\theta_f'(\omega)=\theta_{-f}'(\omega)$, we need to consider only the case $c=1$.

Given $\omega_p\in(0,\pi)$, let $\theta_i:=\theta_{f_i}(\omega_p)$. One can readily verify that 
\begin{align*}
\hspace{-0.5cm}
\sin\theta_i=\frac{(a_i^{2}-1)\sin\omega_p}{|\mathrm{e}^{j\omega_p}+a_i|^2} \quad \text{and} \quad
\theta_{f_i}'(\omega_p)=\frac{a_i^2-1}{|\mathrm{e}^{j\omega_p}+a_i|^2}.
\end{align*}
Since $|a_i|<1$ and $\sin\omega_p >0$, we have $\theta_i\in (-\pi,0)$, $i=1,\cdots,n$. Moreover,
\begin{align*}
\theta_f'(\omega_p) = \sum_{i=1}^n \theta_{f_i}'(\omega_p)
=\sum_{i=1}^n\frac{\sin\theta_i}{\sin\omega_p}.
\end{align*}
Thus, we have
\begin{align*}
\hspace{-5mm}
\theta_f'(\omega_p)
\le \frac{-|\sin(\theta_1+\theta_2+\cdots+\theta_n)|}{\sin\omega_p}
=-\left|\frac{\sin(\theta_{f}(\omega_p))}{\sin\omega_p}\right|,
\end{align*}
where the inequality is shown in~\cite[Lemma 7]{hara2023exact},
and the last equality follows from $\theta_{f}(j\omega_p) = \sum_{i=1}^n \theta_i$. 

For $\omega_p\in(-\pi,0)$, notice that $\theta_f'(\omega_p) = \theta_f'(-\omega_p)$. To see
this, we note that 
\begin{align*}
\theta_f'(\omega)=\mathrm{Im}\left(j\cdot \frac{f'(\ew)\ew}{f(\ew)}\right):=
\mathrm{Im}\left(p_\omega\right),
\end{align*}
and therefore
\begin{align*}
\hspace{-0.5cm}
\theta_f'(-\omega)=\mathrm{Im}\left(j\cdot \frac{f'(\mathrm{e}^{-j\omega})\mathrm{e}^{-j\omega}}{f(\mathrm{e}^{-j\omega})}\right)
=\mathrm{Im}(-\overline{p_{\omega}})=\theta_f'(\omega).
\end{align*}
Since $\sin\left(\theta_f(-\omega_p)\right) = \sin\left(-\theta_f(\omega_p)\right) = -\sin\theta_f(\omega_p)$
and $\sin(-\omega_p)=-\sin\omega_p$, we see that inequality~\eqref{eq:RateChangeReal_p1} remains true for $\omega_p\in(-\pi,0)$.
Finally, it is clear that  inequality~\eqref{eq:RateChangeReal_p1} becomes equality if $f$ is of first order, 
and $\theta_f'(\omega)=0=\sin(\theta_f(\omega))$, $\forall\omega\in[-\pi,\pi]$ if $f$ is of zeroth order.

\section{Proof of Lemma~\ref{lem:c<r}}
\label{appx:lem:c<r}
Without loss of generality, let $f_r(z):=\frac{\alpha_r z^2+\beta_r z+1}{z^2+\beta_r z+\alpha_r}$ and 
$f_c(z):=\frac{\alpha_c z^2+\beta_c z+1}{z^2+\beta_c z+\alpha_c}$. Note that $f_r$ is stable
with two real poles if and only if $\alpha_r$ and $\beta_r$ satisfy the following conditions:
\begin{align}
|\alpha_r|<1, |\beta_r|<\alpha_r+1, \mbox{ and } \beta_r^2\ge 4\alpha_r. \label{cond:real}    
\end{align}
On the other hand, the equivalent conditions for $f_c$ to be stable with complex poles are
\begin{align}
0<\alpha_c<1 \mbox{ and } \beta_c^2<4\alpha_c. \label{cond:complex}
\end{align}
In the following, we will show that, for given $\omega_p\notin\{0,\pm\pi\}$ and $(\alpha_c,\beta_c)$ satisfying
conditions in~\eqref{cond:complex}, one can always find $(\alpha_r,\beta_r)$ such
that \eqref{cond:real} is satisfied, $\theta_{f_r}(\omega_p)=\theta_{f_c}(\omega_p)$ and $\theta_{f_r}'(\omega_p)>\theta_{f_c}'(\omega_p)$. 

Note that $|f_r(\ew)|=|f_c(\ew)|=1$ for all $\omega$ regardless of the values of $\alpha_r, \beta_r, \alpha_c, \beta_c$. 
Therefore, the phase constraint $\theta_{f_r}(\omega_p)=\theta_{f_c}(\omega_p)$ is equivalent to 
$f_r(\mathrm{e}^{j\omega_p})=f_c(\mathrm{e}^{j\omega_p})$, which in turn is equivalent to 
\begin{align}
\hspace{-0.7cm}
(\alpha_r-1)(\beta_c+2\cos\omega_p)=(\alpha_c-1)(\beta_r+2\cos\omega_p).
\label{cond:phase0}
\end{align}
For~\eqref{cond:phase0} to hold, $\alpha_r, \beta_r, \alpha_c, \beta_c$ must satisfy 
\begin{align}
\frac{\alpha_r-1}{\alpha_c-1} = \frac{\beta_r+2\cos\omega_p}{\beta_c+2\cos\omega_p}=\lambda
\label{cond:phase1}
\end{align}
for some $\lambda\in\mathbb{R}$, of which the value will be determined later. 

For a generic stable 2nd-order all-pass function of the form $f(z):=\frac{\alpha z^2+\beta z+1}{z^2+\beta z+\alpha}$, 
one can readily verify that 
\begin{align*}
\sin\theta_f(\omega_p) &= \frac{(\alpha^2-1)\sin2\omega_p+2\beta(\alpha-1)\sin\omega_p}
{|\mathrm{e}^{j2\omega_p}+\beta\mathrm{e}^{j\omega_p}+\alpha|^2}\\
\theta_f'(\omega_p) &= \frac{2(\alpha^2-1)+2(\alpha-1)\beta\cos\omega_p}
{|\mathrm{e}^{j2\omega_p}+\beta\mathrm{e}^{j\omega_p}+\alpha|^2},
\end{align*}
by which
\begin{align}
\label{eq:pcr}
\hspace{-0.6cm}
\theta_f'(\omega_p)=\frac{\sin\theta_f(\omega_p)\cos\omega_p}{\sin\omega_p}+\frac{2(\sin\omega_p)^2(\alpha^2-1)}{D(\omega_p)},
\end{align}
where $D(\omega):=|\mathrm{e}^{j2\omega}+\beta\mathrm{e}^{j\omega}+\alpha|^2$. Let $D_r(\cdot)$ and $D_c(\cdot)$ denote
the $D(\cdot)$ functions with coefficients $(\alpha_r,\beta_r)$ and $(\alpha_c,\beta_c)$, respectively. By the equaliity
$\sin\theta_f(\omega_p)=\sin\theta_c(\omega_p)$, one can verify that 
\begin{align}
D_r(\omega_p)=\lambda^2D_c(\omega_p). 
\label{eq:den}    
\end{align}
By~\eqref{eq:pcr},
we have that $\theta_{f_r}'(\omega_p)>\theta_{f_c}'(\omega_p)$ if and only if 
\begin{align*}
\frac{\alpha_r^2-1}{D_r(\omega_p)}-\frac{\alpha_c^2-1}{D_c(\omega_p)} > 0 \ \Leftrightarrow \ 
\frac{2(\alpha_c-1)(1-\lambda)}{\lambda D_c(\omega_p)} > 0,
\end{align*}
where~\eqref{cond:phase1} and~\eqref{eq:den} have been used. By~\eqref{cond:complex}, we see that 
the aforementioned inequality holds if and only if $\lambda>1$. The remaining proof will show that 
it is possible to find an $\lambda>1$, such that $\alpha_r:=1+\lambda(\alpha_c-1)$ and $\beta_r:=-2\cos\omega_p+\lambda(\beta_c+2\cos\omega_p)$
give rise to a second-order $f_r$ that is stable and has real poles. This requires $(\alpha_r,\beta_r)$ to satisfy~\eqref{cond:real}.
The constraint $|\alpha_r|<1$ leads to
\begin{align}
\label{ineq:lambda1}
0<\lambda<2/(1-\alpha_c):=u_1.
\end{align}
On the other hand, the constraint $|\beta_r|<\alpha_r+1$ gives rise to
\begin{align}
\begin{split}
&\hspace{-0.3cm}\lambda(\beta_c+2\cos\omega_p+\alpha_c-1)>2\cos\omega_p-2,\ \mbox{and}  \\
&\hspace{-0.3cm}\lambda(\beta_c+2\cos\omega_p-\alpha_c+1)<2\cos\omega_p+2   
\end{split}\label{ineq:lambda2}
\end{align}
Since $0<\alpha_c<1$, there are three possible scenarios: 
$\beta_c+2\cos\omega_p+\alpha_c-1<\beta_c+2\cos\omega_p-\alpha_c+1\le 0$;
$0\le \beta_c+2\cos\omega_p+\alpha_c-1<\beta_c+2\cos\omega_p-\alpha_c+1$;
$\beta_c+2\cos\omega_p+\alpha_c-1<0<\beta_c+2\cos\omega_p-\alpha_c+1$.
In all cases,~\eqref{ineq:lambda2} leads to two upper bounds on $\lambda$:
\begin{align}
\begin{cases}
\lambda< u_2, & \mbox{if $\beta_c+2\cos\omega_p>\alpha_c-1$} \\
\lambda< u_3, & \mbox{if $\beta_c+2\cos\omega_p<1-\alpha_c$},
\end{cases}\label{ineq:lambda3}
\end{align}
where $u_2:=(2\cos\omega_p+2)/(\beta_c+2\cos\omega_p-\alpha_c+1)$
and $u_3:=(2\cos\omega_p-2)/(\beta_c+2\cos\omega_p+\alpha_c-1)$. Note that 
the inequalities in~\eqref{ineq:lambda3} imply that $u_2>0$ and $u_3>0$. Now
combining~\eqref{ineq:lambda1} and~\eqref{ineq:lambda3}, it is required that $\lambda<\min\{u_1,u_2,u_3\}$. 
It remains to show that $\min\{u_1,u_2,u_3\}>1$ and there exists an $\lambda\in(1,\min\{u_1,u_2,u_3\})$ such 
that $\beta_r^2-4\alpha_r\ge 0$. 

To see $\min\{u_1,u_2,u_3\}>1$, note that $u_1>2$ since $0<1-\alpha_c<1$. For $u_2$, observe that
\begin{align*}
0&<\beta_c+2\cos\omega_p-\alpha_c+1\\
 &<2\cos\omega_p-(\alpha_c-2\sqrt{\alpha_c}+1)+2\\
 &=2\cos\omega_p+2-(\sqrt{\alpha_c}-1)^2 < 2\cos\omega_p+2,
\end{align*}
which implies $u_2>1$. The arguments for $u_3>1$ are similar, where one may use $-2\sqrt{\alpha_c}<\beta_c$ to show 
$2\cos\omega_p-2<\beta_c+2\cos\omega_p+\alpha_c-1<0$. Lastly, setting $\lambda$ to one of $u_1, \ u_2, \ u_3$, 
we have $\beta_r^2-4\alpha_r\ge 0$. To see this, note that $\alpha_r=-1$ when $\lambda=u_1$, and therefore 
the inequality holds strictly. Moreover, when $\lambda=u_2$ or $u_3$, we have 
$\beta_r^2-4\alpha_r=\lambda^2(\alpha_c-1)^2>0$. Thus, we have shown that when $\lambda=\min\{u_1,u_2,u_3\}>1$,
$\beta_r^2$ is strictly larger than $4\alpha_r$. By continuity, there must exist $\lambda^{\dagger}> 1$
such that the inequality $\beta_r^2\ge 4\alpha_r$ holds for any $\lambda$ satisfying 
$1<\lambda^{\dagger}\le \lambda < \min\{u_1,u_2,u_3\}$. This concludes the proof.

\section{Proof of Lemma~\ref{mp<1st}} 
\label{appx:mp<1st}
To prove Lemma~\ref{mp<1st}, we require the following gain-phase relationship for a 
discrete-time real-rational transfer function $x$ whose poles and zeros are in $\overline{\D}$. A proof can be found in, 
for example,~\cite{Haykin1972}. The proof is based on applying the Cauchy integral theorem to $\log x(z^{-1})$ as $z^{-1}$ traverses the unit circle,  
while noting that $\log x(z^{-1})$ has all its poles outside the unit circle:
\begin{align}
\label{eq:DTgain/phase}
 \theta_x(\omega_p)=\frac{-\sin\omega_p}{\pi}\int_{0}^{\pi} 
 \frac{A_x(\omega)-A_x(\omega_p)}{\cos\omega-\cos\omega_p}d\omega,
\end{align}
where $\omega_p\in[0,\pi]$, $A_x(\omega):=\ln|x(\enw)|$ and $\theta_x(\omega)$ is the phase of $x(\enw)$. 

Now let $f$ be a minimum-phase function from $\RHinf$ and, without loss of generality, assume $\|f\|_{\Hinf}=1$. 
Define $h(z):=\frac{1-z^2}{2z^2}\frac{f'(z)}{f(z)}=\frac{1-z^2}{2z^2}\frac{d}{dz}\log f(z)$. 
Since $f$ is minimum-phase and an element of $\RHinf$, we see that $h$ has all poles in $\D$. 
Moreover, it can be readily verified that 
$h(\enw)=\sin\omega\left(A_f'(\omega)+j\theta_f'(\omega)\right)$. Applying the same derivations in~\cite{Haykin1972} to $h(z^{-1})$, 
the real and imaginary parts of $h(\enw)$ can be related by~\eqref{eq:DTgain/phase} as follows:
\begin{align}
\hspace{-0.5cm}
\begin{split}
&\theta_f'(\omega_p)=\frac{-1}{\pi}\int_{0}^{\pi}\frac{A_f'(\omega)\sin\omega}{\cos\omega-\cos\omega_p}d\omega \\
&= \frac{-1}{\pi}\lim_{\epsilon\to 0}\left[\int_{0}^{\omega_p-\epsilon}\frac{A_f'(\omega)\sin\omega}{\cos\omega-\cos\omega_p}d\omega  \right.\\
&\hspace{2.5cm} \left.+ \int_{\omega_p+\epsilon}^{\pi}\frac{A_f'(\omega)\sin\omega}{\cos\omega-\cos\omega_p}d\omega \right],
\end{split}\label{eq:form4pcr}
\end{align}
where in the first equality, we have used the fact that $\omega_p$ is a peak-gain frequency and therefore $A_f'(\omega_p)=0$.
Now using integration by parts, we have 
\begin{align*}
&\int_{a}^{b}\frac{A_f'(\omega)\sin\omega}{\cos\omega-\cos\omega_p}d\omega 
=\frac{A_f(\omega)\sin\omega}{\cos\omega-\cos\omega_p}\bigg|_a^b\\
&\hspace{0.75cm}-\int_{a}^{b}\frac{(A_f(\omega)-A_f(\omega_p))(1-\cos\omega\cos\omega_p)}{(\cos\omega-\cos\omega_p)^2}d\omega,
\end{align*}
where we used $A_f(\omega_p)=\log|f(j\omega_p)|=\log\|f\|_{h_\infty}=0$. Applying the formula above, it can be verified
that~\eqref{eq:form4pcr} is equal to 
\begin{align}
&\hspace{-0.7cm}
\lim_{\epsilon\to 0}\left[\frac{A_f(\omega_p+\epsilon)\sin(\omega_p+\epsilon)}{\pi(\cos(\omega_p+\epsilon)-\cos\omega_p)}
-\frac{A_f(\omega_p-\epsilon)\sin(\omega_p-
\epsilon)}{\pi(\cos(\omega_p-\epsilon)-\cos\omega_p)}\right]\notag \\
&\hspace{-0.7cm}
+\frac{1}{\pi}\int_0^{\pi}\frac{(A_f(\omega)-A_f(\omega_p))(1-\cos\omega\cos\omega_p)}{(\cos\omega-\cos\omega_p)^2}d\omega.
\end{align}
Moreover, 
\begin{align*}
&\hspace{-0.7cm}
\frac{A_f(\omega_p+\epsilon)\sin(\omega_p+\epsilon)}{(\cos(\omega_p+\epsilon)-\cos\omega_p)}-\frac{A_f(\omega_p-\epsilon)\sin(\omega_p-
\epsilon)}{(\cos(\omega_p-\epsilon)-\cos\omega_p)}\\
&\hspace{-0.7cm}=\frac{\cos\frac{\epsilon}{2}}{2\sin\frac{\epsilon}{2}}(-A_f(\omega_p+\epsilon)+A_f(\omega_p-\epsilon))\\
&\hspace{-0.4cm}+\frac{A_f(\omega_p+\epsilon)\cos(\omega_p+\frac{\epsilon}{2})}{2\sin(\omega_p+\frac{\epsilon}{2})}
-\frac{A_f(\omega_p-\epsilon)\cos(\omega_p-\frac{\epsilon}{2})}{2\sin(\omega_p-\frac{\epsilon}{2})}.
\end{align*}
As $\epsilon\to 0$, clearly the second and the third terms cancel out, and by the L'H\^{o}pital's rule we have
\begin{align*}
&\lim_{\epsilon\to 0}\frac{\cos\frac{\epsilon}{2}}{2\sin\frac{\epsilon}{2}}(-A_f(\omega_p+\epsilon)+A_f(\omega_p-\epsilon))\\
=&\lim_{\epsilon\to 0}\frac{\sin\frac{\epsilon}{2}}{2\cos\frac{\epsilon}{2}}(A'_f(\omega_p+\epsilon)+A'_f(\omega_p-\epsilon))=0
\end{align*}
Thus, we obtain that
\begin{align*}
\hspace{-0.7cm}
\theta_f'(\omega_p)&=\frac{1}{\pi}\int_0^{\pi}\frac{(A_f(\omega)-A_f(\omega_p))(1-\cos\omega\cos\omega_p)}{(\cos\omega-\cos\omega_p)^2}d\omega \\
=-&\frac{1}{\pi}\int_0^{\pi}\frac{|A_f(\omega)-A_f(\omega_p)|(1-\cos\omega\cos\omega_p)}{(\cos\omega-\cos\omega_p)^2}d\omega  \\
=-&\frac{1}{\pi}\int_0^{\pi}\left|\frac{A_f(\omega)-A_f(\omega_p)}{\cos\omega-\cos\omega_p}\right|
\left|\frac{1-\cos\omega\cos\omega_p}{\cos\omega-\cos\omega_p}\right|d\omega,
\end{align*}
where we used the fact that $A_f(\omega)\le A_f(\omega_p)$ for all $\omega\in[0,\pi]$ and the other terms are all positive to obtain the second equality. By the above expression of $\theta_f'(\omega)$, it is clear that $\theta_f'(\omega)\le 0$ regardless of the value of 
$\omega_p$. Now, given $\omega_p\notin\{0,\pm\pi\}$, since $|1-\cos\omega\cos\omega_p|/|\cos\omega-\cos\omega_p|\ge 1$ for all $\omega$, we have 
\begin{align*}
\hspace{-0.7cm}
|\sin\omega_p|\theta_f'(\omega_p)
\le -&\frac{|\sin\omega_p|}{\pi}\int_0^{\pi}\left|\frac{A_f(\omega)-A_f(\omega_p)}{\cos\omega-\cos\omega_p}\right|d\omega\\
&\hspace{-3cm}\le-\frac{|\sin\omega_p|}{\pi}\left|\int_0^{\pi}\frac{A_f(\omega)-A_f(\omega_p)}{\cos\omega-\cos\omega_p}d\omega\right|
=-|\theta_f(\omega_p)|,
\end{align*}
and thus 
\begin{align*}
\theta_f'(\omega_p) \le -\left|\frac{\theta_f(\omega_p)}{\sin\omega_p}\right|.
\end{align*}
This concludes the proof. 

\end{document}